# Guided Wave Propagation in Left-Handed Microstrip Structure Using Dispersive SRR Metamaterial

Clifford M. Krowne*

Microwave Technology Branch, Electronics Technology Division, Naval Research Laboratory, Washington, D.C. 20375-5347

## Abstract

SRR metamaterial is used as a substrate material in a microstrip guided wave structure to determine what the effect is of a material with potentially excessive dispersion or loss or both. A Green's function method readily incorporates the metamaterial permittivity and permeability tensor characteristics. Ab initio calculations are performed to obtain the dispersion diagrams of several complex propagation constant modes of the structure. Analytical analysis is done for the design and interpretation of the results, which demonstrate remarkable potential for realistic use in high frequency electronics while using the LHM for possible field reconfigurations.

It has been shown that very unusual field patterns occur for guided propagating waves in microstrip left-handed material (LHM) structures compatible with microwave integrated circuit technology [1]. Like for focusing, the most arbitrary control of the field pattern obtains when the substrate material can be isotropic, and then modified from isotropy to anisotropy to enhance certain features [2] if desired. For lenses, lack of isotropy can be disastrous, leading to serious wave distortion, and this holds true when studying LHM lenses. For the above reasons, only isotropic substrates will be analyzed in what follows. One of the looming major questions remaining to be answered using LHMs, is what effect does the substrate loss or dispersion have on the characteristics of guided wave propagation while reconfiguring the electromagnetic fields? For low loss or dispersion, or non-dispersive LHMs, one can examine the dispersion diagram in [1] which was used to extract isolated eigenvalue points for making distribution plots. Of course, the use of such ideal crystalline substrates to make non-dispersive devices is the goal, as in the negative refractive (NR) heterostructure bicrystal device creating field asymmetry [3] (Such negative refraction without negative effective index has been studied in dielectric [4] and metal [5] photonic crystals in contrast to the more intensively studied cases of NR in LHM in dielectric photonic crystals [6]). However, significant dispersion may be present, even if the loss is low, in substrates fabricated from ordinary dielectric host right-handed materials (RHMs) with dielectric RHM inclusions, as in photonic crystals [7]. Finally, the use of metallic inclusions, such as split ring-rod combinations (SRRs) [8], may be used, which may have very sizable loss and dispersion.

In this Letter, it is shown, contrary to prevailing wisdom in the microwave community about the use of metallic inclusions and supported by the substantial losses seen in measurements on SRR prisms and other structures in the physics community [9]-[19], that even with extremely dispersive metamaterials, with potential for huge losses, a frequency band (or bands) may be found where the propagation is predominantly lossloss, the wave behavior is backward (left-handed guided wave of reduced dimensionality), and the slow wave phase characteristic comparable to a RHM. This is nothing short of remarkable, and below we will design the metamaterial, highlighting its physics, and selecting reasonable values with which to do realistic simulations. The low loss bands are a result of the 3D guided wave problem being reduced to a single 1D propagation direction, which apparently can be optimized, whereas the 3D lens focusing problem requires a multiplicity of propagation directions, all acting in 3D, which makes the acquisition of low loss propagation much harder.

Effective permittivity $\varepsilon(\omega)$ [20] and permeability $\mu(\omega)$ [21] of the metamaterial can be represented by

$$\varepsilon(\omega) = 1 - \frac{\omega_{pe}^2 - \omega_{oe}^2}{\omega^2 - \omega_{oe}^2 + i\omega\Gamma_e} \quad (1)$$

$$\mu(\omega) = 1 - \frac{F\omega^2}{\omega^2 - \omega_{om}^2 + i\omega\Gamma_m} \quad (2)$$

Here $\omega_{pe}$, $\omega_{oe}$ and $\omega_{om}$ are the effective plasma, and electric and magnetic resonance radian frequencies. $\Gamma_e$ and $\Gamma_m$ are the loss widths in sec$^{-1}$. Real part $\varepsilon_r(\omega)$ of $\varepsilon(\omega)$ is

$$\varepsilon_r(\omega) = \frac{\left(\omega^2 - \omega_{pe}^2\right)\left(\omega^2 - \omega_{oe}^2\right) + \omega^2 \Gamma_e^2}{\left(\omega^2 - \omega_{oe}^2\right)^2 + \omega^2 \Gamma_e^2} \quad (3)$$

As long as we are not too close to resonance, measured against the line width $\Gamma_e$, i.e., $|\omega^2 - \omega_{oe}^2| \gg \omega \Gamma_e$, (3) reduces to

$$\varepsilon_r(\omega) = 1 - \frac{\omega_{pe}^2 - \omega_{oe}^2}{\omega^2 - \omega_{oe}^2} = 1 - \frac{\omega_{pe}^2}{\omega^2} \quad (4)$$

the last equality arising from having continuous rods making $\omega_{oe} = 0$. For a desired $\varepsilon_r(\omega)$ at a specific frequency, (3) may be inverted to yield

$$\omega_{pe} = \sqrt{\left(\omega^2 + \Gamma_e^2\right)\left[1 - \varepsilon_r(\omega)\right]}$$

$$\approx \omega\sqrt{1 - \varepsilon_r(\omega)} \quad (5)$$

So if we wish to have $\varepsilon_r(\omega) = -2.5$ at f = 80 GHz, then $\omega_{pe} = \omega\sqrt{3.5} = 150$ GHz by evaluating (5).

Imaginary part $\varepsilon_i(\omega)$ of $\varepsilon(\omega)$ is

$$\varepsilon_i(\omega) = \frac{\omega \Gamma_e \left(\omega_{pe}^2 - \omega_{oe}^2\right)}{\left(\omega^2 - \omega_{oe}^2\right)^2 + \omega^2 \Gamma_e^2} \quad (6)$$

Again for $\omega_{oe} = 0$, (6) becomes

$$\varepsilon_i(\omega) = \frac{\Gamma_e \omega_{pe}^2}{\omega\left(\omega^2 + \Gamma_e^2\right)} \approx \frac{\Gamma_e}{\omega}\left(\frac{\omega_{pe}}{\omega}\right)^2 \quad (7)$$

Since $\omega_{pe}$ has independence from rod metal electron carrier density, given by

$$\omega_{pe}^2 = \frac{2\pi c_0^2}{a^2 \ln(a/r_e)} \quad (8)$$

for a chosen lattice spacing a to wire radius $r_e$ ratio, $a/r_e$, a can be solved for in (8)

$$a^2 = \frac{2\pi c_0^2}{\omega_{pe}^2 \ln(a/r_e)} \quad (9)$$

Setting $a/r_e$ = 109.3, inserting the free space light velocity $c_0$ and $\omega_{pe}$, we find that a = 0.690 mm corresponding to $r_e$ = 6.313 μm. With this same lattice spacing to wire radius ratio, one can determine $\Gamma_e$ as

$$\Gamma_e = \frac{\varepsilon_0}{\pi}\left(\frac{a}{r_e}\right)^2 \frac{\omega_{pe}^2}{\sigma_r} \quad (10)$$

One notices that that the linewidth does indeed depend on the carrier density through metallic rod conductivity $\sigma_r$, and this is where potentially one can suffer tremendous ohmic losses, the bane of physicists and engineers trying either to reduce SRR based structure losses or circuit losses. If we use aluminum as had Pendry, taking $\sigma_r = \sigma_{Al}$ = $3.65 \times 10^7$ $\Omega^{-1}$/m, $\Gamma_e$ = 0.1305 GHz. Once $\Gamma_e$ is known, by (7) the frequency variation of the imaginary part of the permittivity is fixed.

Turning our attention to the effective permeability, real part $\mu_r(\omega)$ of $\mu(\omega)$ is

$$\mu_r(\omega) = \frac{(\omega^2 - \omega_{om}^2 - F\omega^2)(\omega^2 - \omega_{om}^2) + \omega^2\Gamma_m^2}{(\omega^2 - \omega_{om}^2)^2 + \omega^2\Gamma_m^2} \quad (11)$$

For a desired $\mu_r(\omega)$ at a specific frequency, (11) produces an equation for the magnetic resonance frequency,

$$(\omega^2 - \omega_{om}^2)^2(\mu_r - 1) + F\omega^2(\omega^2 - \omega_{om}^2) + \omega^2\Gamma_m^2(\mu_r - 1) = 0 \quad (12)$$

This is a quadratic equation in $\omega^2 - \omega_{om}^2$ and so will generate a solution for $\omega_{om}$ once the correct root is determined since the squared root is displaced from the desired squared resonance value. As long as we are not too close to resonance, measured against the line width $\Gamma_m$, i.e., $|\omega^2 - \omega_{om}^2| \gg \omega\Gamma_m$, we may reduce (11) directly with the result

$$\mu_r(\omega) = 1 - \frac{F}{\omega^2 - \omega_{om}^2} \tag{13}$$

For a desired $\mu_r(\omega)$ at a specific frequency, (13) may be inverted to yield the magnetic resonance frequency

$$\omega_{om} = \omega\sqrt{1 - \frac{F}{1-\mu_r(\omega)}} \tag{14}$$

So if we wish to have $\mu_r(\omega) = -2.5$ at f = 80 GHz, then $\omega_{om} = \omega\sqrt{0.8564} = 74.03$ GHz by evaluating (14), using a value of F=0.5027 obtained from using the ratio of cylinder radius to lattice spacing,

$$F = \pi\left(\frac{r_m}{a}\right)^2 \tag{15}$$

The F value quoted makes $r_m$ 40 % of a, i.e., $r_m/a = 2/5$, giving $r_m = 0.2761$ mm. For the sheets split on cylinders model, effective magnetic resonance frequency

$$\omega_{om} = \sqrt{\frac{3dc_0^2}{\pi^2 r_m^3}} \tag{16}$$

can be inverted to find concentric cylinder spacing d,

$$d = \frac{\pi^2 \omega_{om}^2 r_m^3}{3c_0^2} \tag{17}$$

Inserting in the values already calculated, we find that d = 0.1668 mm. This model is like the SRR in that d will be like the gap in the split ring, and the ring radius is like the cylinder radius. Clearly the gap in the split ring has a capacitance dependent on the specific edge cross-section presented to the gap, whereas very short height cylinders look like rings with overlapping surfaces having length $2\pi r_m$, and one only needs to adjust the parameters to make one model equivalent to the other.

Imaginary part $\mu_i(\omega)$ of $\mu(\omega)$ is

$$\mu_i(\omega) = \frac{\omega^3 \Gamma_m F}{\left(\omega^2 - \omega_{om}^2\right)^2 + \omega^2 \Gamma_m^2} \approx \frac{\omega^3 \Gamma_m F}{\left(\omega^2 - \omega_{om}^2\right)^2} \tag{18}$$

One can specify linewidth which is required to get the exact frequency of $\mu_i(\omega)$ using

$$\Gamma_m = \frac{2 R_s}{r_m \mu_0} \tag{19}$$

where $R_s$ is the surface resistance of the metal cylinders to electromagnetic waves [22], [23]. It is an inverse function of metallic conductivity of the cylinders $\sigma_c$, allowing (19) to be written as

$$\Gamma_m = \frac{1}{r_m} \sqrt{\frac{2\omega}{\sigma_c \mu_0}} \tag{20}$$

Again using the conductivity of aluminum, $\sigma_c = \sigma_{Al}$, the magnetic linewidth is found to be $\Gamma_m = 0.0853$ GHz.

The guiding structure to be simulated (Fig. 1) has a substrate thickness $h_s = 0.5$ mm, an air region thickness $h_a = 5$ mm, vertical perfectly conducting walls separation $B = 2b = 5$ mm and a microstrip metal width $w = 0.5$ mm, as in [1]. Certainly the lateral dimension of the structure B is substantially greater than the lattice spacing (B/a = 7.25), one requirement for the metamaterial to be used in our simulation. The longitudinal direction, for a uniform guiding structure, as this is, is taken as infinite, and automatically satisfies the metamaterial size requirement (i.e., L >> a, $L \to \infty$). However, the substrate thickness is on the order of lattice spacing, which is not ideal, but we nevertheless accept it for purposes of needing some roughly useable value in order to perform simulations in what follows. Finally, the electromagnetic wavelength to lattice spacing ratio λ/a = 5.43, is substantial. Because of the logarithmic relationship entailed in solving for $r_e$ in (9) once a has been selected, a noticeable reduction in a by small integer factors can have drastic affects on $r_e$ , reducing by many orders of magnitude its size, moving from the μm range to the nm range, something still possible using conductive carbon nanotubes.

In the above calculations we have been careful not to assume the metals of the split rings (or cylinders) are the same, $\sigma_c \neq \sigma_r$ . In fact, one may be made out of gold, and the other out of aluminum, platinum, silver, or copper to name a few common metals. Range of conductivities of these metals is from $1.02 \times 10^7$ $\Omega^{-1}$/m to $6.17 \times 10^7$ $\Omega^{-1}$/m [20].

Also of interest for the effective permittivities and permeabilities, are their crossover frequencies. From (3), the electric crossover frequency $\omega_{eco}$ which satisfies $\varepsilon_r(\omega) = 0$, obeys

$$\omega^4 - \omega^2\left(\omega_{pe}^2 + \omega_{oe}^2 + \Gamma_e^2\right) + \omega_{pe}^2 \omega_{oe}^2 = 0 \tag{21}$$

When $\omega_{oe} = 0$, which is our present case, it is easier to examine the numerator of (3) directly, than to solve (21), yielding

$$\omega_{eco} = \sqrt{\omega_{pe}^2 - \Gamma_e^2} \approx \omega_{pe} \tag{22}$$

From (11), the magnetic crossover frequency $\omega_{mco}$ which satisfies $\mu_r(\omega) = 0$, obeys

$$\omega^4(1-F) - \omega^2\left[\omega_{om}^2(2+F) + \Gamma_m^2\right] + \omega_{om}^4 = 0 \tag{23}$$

One of the solutions gives a value close to the resonance frequency when the linewidth is small, as we have seen it is for our case. For that limiting case, (23) considerably simplifies, and one may also readily find the solution by working with (11) directly, finding

$$\omega_{mco} = \frac{\omega_{om}}{\sqrt{1-F}} \tag{24}$$

Placing the values already calculated into the right-hand-side of (24), $\omega_{mco}$ = 104.98 GHz.

For the LHM/NPV structure (NPV = negative phase velocity), in the limiting case of $\Gamma_e \to 0$ and $\Gamma_m \to 0$, two $\gamma$ solutions exist which have $\alpha = 0$, and $\bar{\beta} = \beta/k_0 =$ 1.177647 and 1.786090 (quoted values for number of current basis functions $n_x$ and $n_z = 1$ and spectral expansion terms n = 200). One corresponds to a forward wave for non-dispersive intrinsic LHMs ($\bar{\beta} = \beta/k_0 = 1.78609$) where the product of the integrated Poynting vector (net power through the cross–section) and phase vector in the z – direction is $\oint \mathbf{P}_z \bullet \beta\hat{\mathbf{z}} dA > 0$ (dA is the differential cross–sectional element) or equivalently $\mathbf{v}_{gl} \cdot \mathbf{v}_{pl} > 0$. The other solution is a backward wave for non-dispersive intrinsic LHMs ($\bar{\beta} = \beta/k_0 = 1.177647$) where the product of the integrated Poynting vector and the phase vector in the z – direction is $\oint \mathbf{P}_z \bullet \beta\hat{\mathbf{z}} dA < 0$ or equivalently $\mathbf{v}_{gl} \cdot \mathbf{v}_{pl} < 0$. Here $\mathbf{v}_{gl}$ and $\mathbf{v}_{pl}$ are respectively the group and phase longitudinal velocities. These modes are referred to as fundamental modes in the sense that as $\omega \to 0$, a solution exists.

As damping loss is turned on and becomes finite, only the larger guided wave eigenvalue solution will exist, and the smaller guided wave solution will cease to exist. One of the effects of utilizing a highly dispersive SRR like substrate is that it converts the forward wave non-dispersive solution into a backward wave (at least up to 93 GHz, then it becomes a slightly forward wave until 103 GHz). The fundamental mode is shown in Fig. 2 (labeled R1: light blue $\bar{\alpha} = \alpha/k_0$ and green $\bar{\beta} = \beta/k_0$ curves), not extending beyond either 74 GHz on the low end or 105 GHz on the high end because $sg[\varepsilon_r(\omega)]sg[\mu_r(\omega)] = -1$, and that product causes evanescent propagation to occur where $\beta$ is extremely tiny. Below $f_{om}$ and above $f_{mco}$, $\mu_r > 0$ and $\varepsilon_r < 0$. However, between these critical frequency points, $sg[\varepsilon_r(\omega)]sg[\mu_r(\omega)] = +1$ and the wave propagates with $\mu_r < 0$ and $\varepsilon_r < 0$. Out of the range plotted, above the electric cross-over frequency $f_{eco}$ (near $f_{pe}$ for low loss) where effective permittivity $\varepsilon_r = 0$, $sg[\varepsilon_r(\omega)]sg[\mu_r(\omega)] = +1$ will occur again, with $\mu_r > 0$ and $\varepsilon_r > 0$ this time, being completely reversed from the region between $f_{om}$ and $f_{mco}$.

Results for the two other modes R2 (dark blue $\bar{\alpha} = \alpha/k_0$ and orange $\bar{\beta} = \beta/k_0$ curves) and R3 (dark magenta $\bar{\alpha} = \alpha/k_0$ and red $\bar{\beta} = \beta/k_0$ curves) are also shown in Fig. 2. They clearly have much higher attenuation than the fundamental mode. These modes do show forward wave behavior where the curve slopes go positive near 77.7 GHz and persist to 80 GHz.

The dispersion curves for the LHM/NPV substrate are plotted for the modes in Fig. 2 using $n_x = n_z = 5$ and $n = 500$ with frequency increments of $\Delta f = 0.0025$ GHz between eigenvalue points ( only small differences with solutions at $n_x = n_z = 9$ and $n = 900$ and $n_x = n_z = 1$ and $n = 100$ have been found, on the order of less than a few tenths of a percent). At 80 GHz for the fundamental mode, $\bar{\gamma} = (\bar{\alpha}, \bar{\beta}) = (0.0320222, 1.79365)$ for $n_x = n_z = 1$ and $n = 100$, whereas for $n_x = n_z = 9$ and $n = 900$, $\bar{\gamma} = (0.0322745, 1.79116)$. Calculated values used the parameter settings mentioned above at room temperature, but note that $\Gamma_e$ and $\Gamma_m$ could be reduced further using superconductive metals at reduced temperatures. Permittivity and permeability at 80 GHz are $\varepsilon = (-2.5156, 5.77348 \times 10^{-3})$ and $\mu = (-2.4817, 2.5713 \times 10^{-2})$, corresponding to loss tangents ($\varepsilon_i/\varepsilon_r$ or $\mu_i/\mu_r$) of $2.28 \times 10^{-3}$ and $1.04 \times 10^{-2}$. The reason why $\alpha$ rises and $\beta$ falls after this frequency in Fig. 2 is that both $\varepsilon_r$ and $\mu_r$ are becoming ever more positive, providing much less left-handed material advantage, with the relative rates of change of $\varepsilon_r$ and $\mu_r$ determining the details of the curves. Of course, once the frequency $f_{mco}$ is hit, any effective propagation ceases.

Examination of the fundamental mode at 80 GHz shows that the attenuation compared to the phase behavior is roughly two orders of magnitude smaller ($\beta/\alpha = 56$).

This is truly a remarkable results, since by intuition alone, one might have thought that a metamaterial based upon metallic resonant objects would be extremely lossy. And indeed the guided wave structure does present huge losses below 75 GHz which is approaching the magnetic resonance point, while above 80.5 GHz one is entering the region where neither $\varepsilon_r$ nor $\mu_r$ hold their simultaneous substantial negative real parts and the LHM behavior begins to die out. But between these two bounds, a sizable bandwidth exists where $\alpha$ is relatively tiny compared to $\beta$, giving us a freely propagating wave which is allowed by a substrate acting as a left-handed material, and propagating backward. It is easy to see that indeed in this band the wave is backward, since $v_{gl} = d\omega(k)/dk = 1/[dk(\omega)/d\omega] = 1/[dk(f)/df] < 0$, and inspection of the curve shows $dk(f)/df < 0$.

Even the higher order modes R2 and R3 have low loss bands, but they are much smaller than the fundamental mode, existing between 75 GHz and 76.7 GHz (R2) and 77.5 GHz (R3).

($n_x=n_z=1$, n=200) and 2.24623 ($n_x=n_z=5$, n=500) all asymmetric basis set; β(rotated-microstrip) = 2.11530 asymmetric basis and 2.11604 symmetric basis ($n_x=n_z=1$, n=200).

FIGURE CAPTIONS

1. (color) Cross-section of guided wave structure.

2. (color) Dispersion curves for LHM structure.

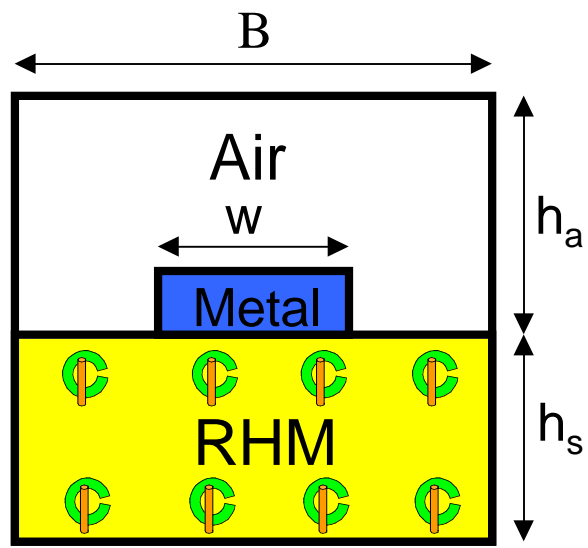

FIGURE 1

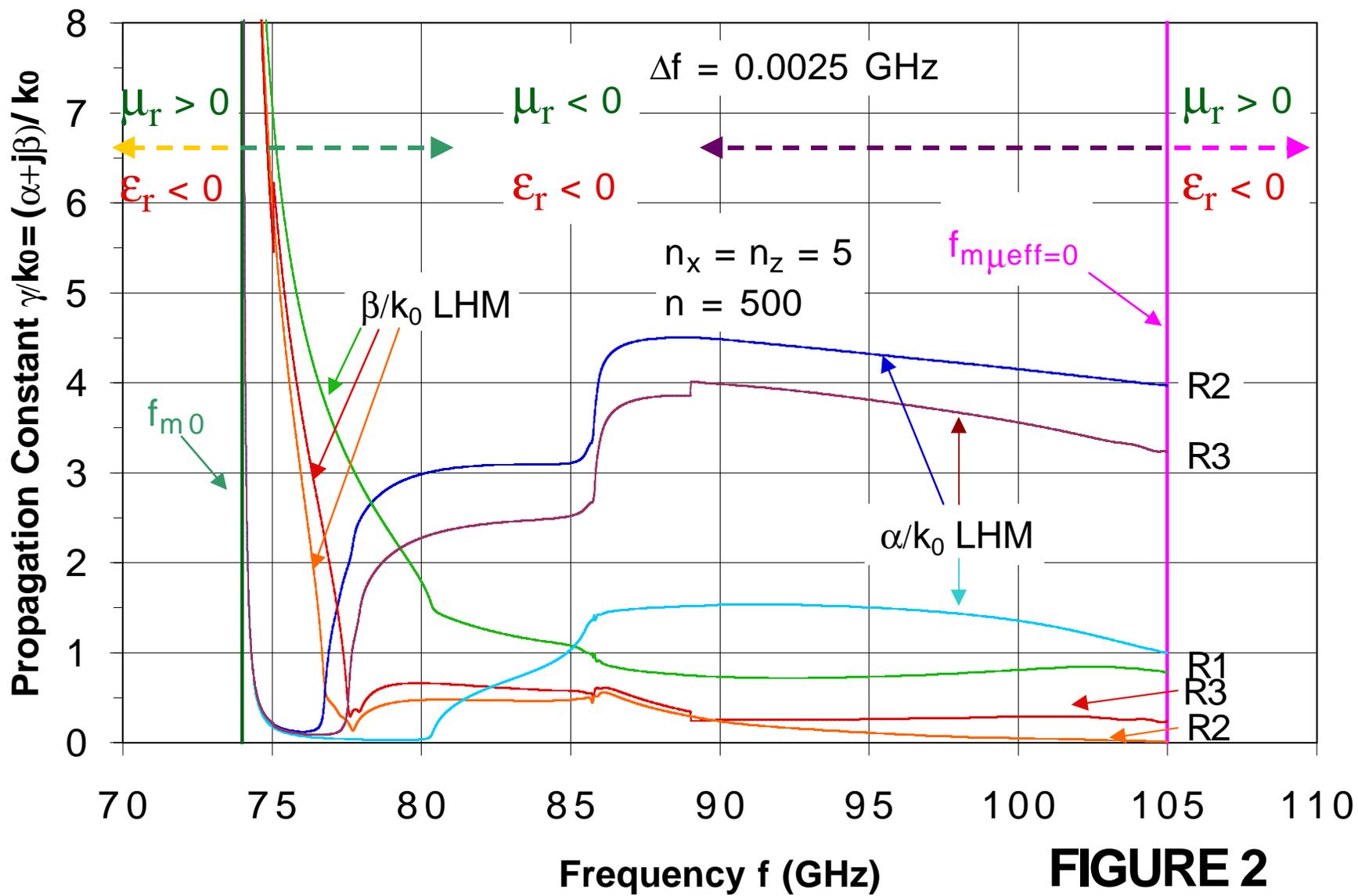

FIGURE 2